\documentclass[12pt]{article}

\usepackage{cancel}
\usepackage{enumitem}
\usepackage{amssymb}
\usepackage{soul}
\usepackage{tabularx}
\usepackage{xcolor}
\usepackage{booktabs}
\usepackage{subcaption} 
\usepackage{colortbl}
\usepackage{authblk}
\usepackage{xspace}
\DeclareUnicodeCharacter{00A0}{ }

\newcolumntype{C}{>{\centering\arraybackslash}X}
\usepackage[T1]{fontenc}
\usepackage{epsfig}
\usepackage{latexsym}
\usepackage{graphicx}
\usepackage{amsmath}
\usepackage{amsfonts}   
\usepackage{amssymb}    
\usepackage{float}
\usepackage{bm}
\usepackage{url}
\usepackage{hyperref} 
\usepackage[nodisplayskipstretch]{setspace}
\usepackage{url}
\def\lsim{\raise0.3ex\hbox{$\;<$\kern-0.75em\raise-1.1ex\hbox{$\sim\;$}}}
\def\gsim{\raise0.3ex\hbox{$\;>$\kern-0.75em\raise-1.1ex\hbox{$\sim\;$}}}

\def    \beq            {\begin{equation}}
\def    \eeq            {\end{equation}}
\def    \bea           {\begin{eqnarray}}
\def    \eea           {\end{eqnarray}}

\def \mn{\mu\nu{\rm SSM}}

\def\g2{{\rm GeV}^2}

\def\sw2{sin^2 \theta_w}

\def\a^tau{\alpha_{\tau}}

\def\beq{\begin{equation}}
\def\eeq{\end{equation}}
\def\beqa{\begin{eqnarray}}
\def\eeqa{\end{eqnarray}}

\newcommand{\tev}{\;\textrm{TeV}}
\newcommand{\gev}{\;\textrm{GeV}}

\newcommand{\newc}{\newcommand}
\newc\BR{BR}
\newc{\akappa}{A_{\kappa} }
\newc\deltagmtwo{\delta (g-2)_{\mu}} 
\newc\deltaamu{\Delta a_{\mu}}

\def\anti{\overline}

\def\la{\lambda}

\def\ka{\kappa}

\newc{\haa}{BR\(h_1\to a_1 a_1\)}
\newc{\abb}{BR\(a_1\to b\anti{b}\)}
\newc{\hbb}{BR\(h_1\to b\anti{b}\)}
\newc{\abund}{\Omega h^2}
\newc\bsgamma{b\rightarrow s \gamma }
\newc\bxsgamma{\overline{B}\rightarrow X_{s}\gamma}
\newc\brbsgamma{\BR(\overline{B}\rightarrow X_s\gamma)}

\newcommand\ReDiag{\mathop{%
  \raise .5pt\hbox{[}%
  \widetilde{\mathrm{Re}}%
  \raise .5pt\hbox{]}}}
\newcommand\ReOffDiag{\mathop{%
  \raise .5pt\hbox{$\llbracket$}%
  \widetilde{\mathrm{Re}}%
  \raise .5pt\hbox{$\rrbracket$}}}

\newcommand{\mnSSM}{\ensuremath{\mu\nu\mathrm{SSM}}}

\def\order#1{\ensuremath{{\cal O}(#1)}}

\def\la{\lambda}

\def\gmin2{\ensuremath{(g-2)_\mu}}

\definecolor{Orange}{named}{orange}
\definecolor{Purple}{named}{purple}
\definecolor{Lightblue}{cmyk}{0.9,0.1,0.1,0.3}
\definecolor{dgelborange}{cmyk}{0.,0.3,0.5, 0.}
\definecolor{Lila}{rgb}{0.5,0.,1}
\definecolor{Darkgreen}{rgb}{0.,.7,0.2}

\allowdisplaybreaks

\usepackage{feynmp}
\graphicspath{{./Figures/Diagrams/}}
\usepackage{enumitem} 

\usepackage{array, makecell}
\usepackage{boldline}
\usepackage[nosort]{cite}

\title{\bf{
Searching for stop LSP at the LHC 
} }

\author[a,b,c]{Essodjolo Kpatcha\thanks{kpatcha.essodjolo@uam.es}}
\author[d]{I\~naki Lara\thanks{inaki.lara@fuw.edu.pl}}
\author[e,f]{Daniel~E.~L\'opez-Fogliani\thanks{daniel.lopez@df.uba.ar}}
\author[a,b]{Carlos~Mu\~noz\thanks{c.munoz@uam.es}} 
\author[g]{Natsumi Nagata\thanks{natsumi@hep-th.phys.s.u-tokyo.ac.jp}}
\author[h]{Hidetoshi Otono\thanks{otono@phys.kyushu-u.ac.jp}}

\affil[a]{Instituto de F\'{\i}sica Te\'{o}rica (IFT) UAM-CSIC, 
  Campus de Cantoblanco, 28049 Madrid, Spain}
\affil[b]{Departamento de F\'{\i}sica Te\'{o}rica, Universidad Aut\'{o}noma de Madrid (UAM),
Campus de Cantoblanco, 28049 Madrid, Spain}

\affil[c]{Université Paris-Saclay, CNRS/IN2P3, IJCLab, 91405 Orsay, France}

 \affil[d] 
  {Faculty of Physics, University of Warsaw, Pasteura 5, 02-093 Warsaw, Poland}
  \affil[e]{Instituto de F\'isica de Buenos Aires UBA \& CONICET, Departamento de F\'isica,
 Facultad de Ciencia Exactas y Naturales, Universidad de Buenos Aires, 
1428 Buenos Aires, Argentina}
\affil[f]{
{Pontificia Universidad Católica Argentina, 
Av. Alicia Moreau de Justo 1500, 
1107~Buenos~Aires, Argentina}}
\affil[g] {Department of Physics, University of Tokyo, 
Tokyo 113-0033, Japan}
\affil[h] {Research Center for Advanced Particle Physics, Kyushu University, Fukuoka 819-0395, Japan 
}

\date{}

\begin{document}

\thispagestyle{empty}
\begin{flushright}
\mbox{}
IFT--UAM/CSIC-21-137\\
FTUAM-21-8\\
KYUSHU-RCAPP-2021-02
\end{flushright}
{\let\newpage\relax\maketitle}

\maketitle

\begin{abstract}
We analyze relevant signals expected at the LHC for a stop as the lightest supersymmetric particle (LSP).
The discussion is carried out in the framework of the $\mu\nu$SSM, where the presence of $R$-parity violating couplings involving right-handed neutrinos solves the $\mu$-problem and reproduces neutrino data.
The stops are pair produced at the LHC, decaying with displaced vertices to a lepton and a bottom quark or a neutrino and a top quark. 
We compare the predictions of this scenario with ATLAS and CMS searches for long-lived particles.
To analyze the parameter space
we sample the $\mu\nu$SSM for a stop LSP
using a likelihood data-driven method, and paying special attention to reproduce the current experimental data on neutrino and Higgs physics, as well as flavor observables.
Our results translate into a lower limit on the mass of the left (right) stop LSP of 1068 GeV (1341 GeV), corresponding to a decay length of 
1.86 mm (6.61 mm).

\end{abstract}



\clearpage 

\tableofcontents 

\section{Introduction}
\label{sec:intro}

The `$\mu$~from~$\nu$' Supersymmetric Standard Model
(\mnSSM)~\cite{LopezFogliani:2005yw} is a highly predictive model (for a recent review, see Ref.~\cite{Lopez-Fogliani:2020gzo}), alternative to the 
Minimal Supersymmetric Standard Model (MSSM)~\cite{Nilles:1983ge,Barbieri:1987xf,Haber:1984rc,Gunion:1984yn, Martin:1997ns}.
The \mnSSM\ solves the
$\mu$-problem and the $\nu$-problem (neutrino masses) simultaneously without
the need to introduce additional energy scales beyond the supersymmetry (SUSY)-breaking scale. In contrast to the MSSM, and to the Next-to-MSSM (NMSSM)~\cite{Maniatis:2009re,Ellwanger:2009dp}, $R$-parity
and lepton number are not conserved,
leading to a completely different
phenomenology characterized by distinct prompt or displaced
decays of the lightest supersymmetric particle (LSP),
producing multi-leptons/jets/photons with small/moderate missing transverse energy (MET) from 
neutrinos~\cite{Ghosh:2017yeh,Lara:2018rwv,Lara:2018zvf,Kpatcha:2019gmq,Kpatcha:2019pve,Heinemeyer:2021opc}.
The smallness of neutrino masses is directly related with the low decay width of the LSP. Actually, it is also related to the existence of possible candidates for decaying dark matter in the model.
This is the case of 
the gravitino~\cite{Choi:2009ng,GomezVargas:2011ph,Albert:2014hwa,GomezVargas:2017,Gomez-Vargas:2019mqk}, or the axino~\cite{Gomez-Vargas:2019vci}, with lifetimes greater than the age of the Universe.
It is also worth mentioning concerning cosmology, that baryon asymmetry might be realized in the
$\mn$ through electroweak (EW) baryogenesis~\cite{Chung:2010cd}. 
The EW sector of the $\mn$ can also explain~\cite{Kpatcha:2019pve,Heinemeyer:2021opc}
 the longstanding
discrepancy between the experimental result for the anomalous magnetic
moment of the muon~\cite{Abi:2021gix,Albahri:2021ixb} and its SM prediction~\cite{Aoyama:2020ynm}.\footnote{
In this work we will not try to explain it
since we are interested in the analysis of a stop LSP through the decoupling of the rest of the SUSY spectrum.}

Because of $R$-parity violation (RPV) in the $\mn$, basically all SUSY particles are candidates for the LSP, and therefore analyses of the LHC phenomenology associated to each candidate are necessary to test them.
This crucial task, given the current experimental results on SUSY searches, has been concentrated  up to now on the EW sector of the $\mn$, in particular analyzing left sneutrinos, the right smuon and the bino as candidates for the LSP~\cite{Ghosh:2017yeh,Lara:2018rwv,Lara:2018zvf,Kpatcha:2019gmq,Kpatcha:2019pve,Heinemeyer:2021opc}.
The aim of this work is to start a systematic analysis of the color sector of the $\mn$.
We will concentrate here on the SUSY partners of the top quark as the LSP candidates, i.e. the left and right stops. 

In particular, we will study the constraints on the parameter space by sampling the model to get either the left stop as the LSP or the right stop as the LSP in a wide range of masses. We will pay special attention
to reproduce 
neutrino masses and mixing angles~\cite{Capozzi:2017ipn,deSalas:2017kay,deSalas:2018bym,Esteban:2018azc}. 
In addition, we will impose on the resulting parameters agreement with Higgs data as well as with flavor observables.

The paper is organized as follows. In Section~\ref{sec:model}, we will review the $\mn$ and its relevant parameters for our analysis of neutrino, neutral Higgs and stop sectors.
In Section~\ref{sec:stops},
we will introduce the phenomenology of the stop LSP, studying its pair production channels at the LHC and its signals. The latter consist of displaced vertices with 
a lepton and a bottom quark or a neutrino and a top quark.
In Section~\ref{strategy}, we will discuss the strategy that we will employ to
perform scans searching for points of the parameter space of our scenario compatible with current experimental data on neutrino and Higgs physics, as well as flavor observables such as 
$B$ and $\mu$ decays.
The results of these scans will be presented 
in Section~\ref{sec:results}, and applied to show the 
current reach of the LHC search on the parameter space of the stop LSP based on ATLAS and CMS 
results~\cite{ATLAS:2019gqq,CMS:2017poa,CMS:2020iwv,CMS:2021kdm,ATLAS:2020xyo,ATLAS:2017tny,Proceedings:2018het,CMS:2018ncu,CMS:2018lab,ATLAS:2021jyv,ATLAS:2017drc,ATLAS:2020dsf,CMS:2021eha}.
Finally, 
our conclusions are left for Section~\ref{sec:conclusion}.



\section{The $\mu\nu$SSM }
\label{sec:model}

In the $\mn$~\cite{LopezFogliani:2005yw,Lopez-Fogliani:2020gzo}, the particle content of the MSSM
is extended by RH neutrino superfields $\hat \nu^c_i$. 
{The simplest superpotential of the model 
is the following~\cite{LopezFogliani:2005yw,Escudero:2008jg,Ghosh:2017yeh}: 
\bea
W &=&
\epsilon_{ab} \left(
Y_{e_{ij}}
\, \hat H_d^a\, \hat L^b_i \, \hat e_j^c +
Y_{d_{ij}} 
\, 
\hat H_d^a\, \hat Q^{b}_{i} \, \hat d_{j}^{c} 
+
Y_{u_{ij}} 
\, 
\hat H_u^b\, \hat Q^{a}
\, \hat u_{j}^{c}
\right)
\nonumber\\
&+& 
\epsilon_{ab} \left(
Y_{{\nu}_{ij}} 
\, \hat H_u^b\, \hat L^a_i \, \hat \nu^c_j
-
\lambda_i \, \hat \nu^c_i\, \hat H_u^b \hat H_d^a
\right)
+
\frac{1}{3}
\kappa_{ijk}
\hat \nu^c_i\hat \nu^c_j\hat \nu^c_k,
\label{superpotential}
\eea
where the summation convention is implied on repeated indices, with $i,j,k=1,2,3$ the usual family indices of the SM 
and $a,b=1,2$ $SU(2)_L$ indices with $\epsilon_{ab}$ the totally antisymmetric tensor, $\epsilon_{12}= 1$. 
}

{Working in the framework of a typical low-energy SUSY, the Lagrangian  containing the soft SUSY-breaking terms related to $W$ 
is given by:
\bea
-\mathcal{L}_{\text{soft}}  =&&
\epsilon_{ab} \left(
T_{e_{ij}} \, H_d^a  \, \widetilde L^b_{iL}  \, \widetilde e_{jR}^* +
T_{d_{ij}} \, H_d^a\,   \widetilde Q^b_{iL} \, \widetilde d_{jR}^{*} 
+
T_{u_{ij}} \,  H_u^b \widetilde Q^a_{iL} \widetilde u_{jR}^*
+ \text{h.c.}
\right)
\nonumber \\
&+&
\epsilon_{ab} \left(
T_{{\nu}_{ij}} \, H_u^b \, \widetilde L^a_{iL} \widetilde \nu_{jR}^*
- 
T_{{\lambda}_{i}} \, \widetilde \nu_{iR}^*
\, H_d^a  H_u^b
+ \frac{1}{3} T_{{\kappa}_{ijk}} \, \widetilde \nu_{iR}^*
\widetilde \nu_{jR}^*
\widetilde \nu_{kR}^*
\
+ \text{h.c.}\right)
\nonumber\\
&+&   
m_{\widetilde{Q}_{ijL}}^2
\widetilde{Q}_{iL}^{a*}
\widetilde{Q}^a_{jL}
{+}
m_{\widetilde{u}_{ijR}}^{2}
\widetilde{u}_{iR}^*
\widetilde u_{jR}
+ 
m_{\widetilde{d}_{ijR}}^2
\widetilde{d}_{iR}^*
\widetilde d_{jR}
+
m_{\widetilde{L}_{ijL}}^2
\widetilde{L}_{iL}^{a*}  
\widetilde{L}^a_{jL}
\nonumber\\
&+&
m_{\widetilde{\nu}_{ijR}}^2
\widetilde{\nu}_{iR}^*
\widetilde\nu_{jR} 
+
m_{\widetilde{e}_{ijR}}^2
\widetilde{e}_{iR}^*
\widetilde e_{jR}
+ 
m_{H_d}^2 {H^a_d}^*
H^a_d + m_{H_u}^2 {H^a_u}^*
H^a_u
\nonumber \\
&+&  \frac{1}{2}\, \left(M_3\, {\widetilde g}\, {\widetilde g}
+
M_2\, {\widetilde{W}}\, {\widetilde{W}}
+M_1\, {\widetilde B}^0 \, {\widetilde B}^0 + \text{h.c.} \right).
\label{2:Vsoft}
\eea

In the early universe not only the EW symmetry is broken, but in addition to the neutral components of the Higgs doublet fields $H_d$ and $H_u$ also the left and right sneutrinos $\widetilde\nu_{i}$ and $\widetilde\nu_{iR}$
acquire a vacuum expectation value (VEV). {With the choice of CP conservation, they develop real VEVs denoted by:  
\begin{eqnarray}
\langle H_{d}^0\rangle = \frac{v_{d}}{\sqrt 2},\quad 
\langle H_{u}^0\rangle = \frac{v_{u}}{\sqrt 2},\quad 
\langle \widetilde \nu_{iR}\rangle = \frac{v_{iR}}{\sqrt 2},\quad 
\langle \widetilde \nu_{i}\rangle = \frac{v_{i}}{\sqrt 2}.
\end{eqnarray}
The EW symmetry breaking is induced by the soft SUSY-breaking terms
producing
$v_{iR}\sim {\order{1 \tev}}$ as a consequence of the right sneutrino minimization equations in the scalar potential~\cite{LopezFogliani:2005yw,Escudero:2008jg,Ghosh:2017yeh}.
Since $\widetilde\nu_{iR}$ are gauge-singlet fields,
the $\mu$-problem can be solved in total analogy to the
NMSSM
through the presence in the superpotential (\ref{superpotential}) of the trilinear 
terms $\lambda_{i} \, \hat \nu^c_i\,\hat H_u \hat H_d$.
Then, the value of the effective $\mu$-parameter is given by 
$\mu=
\la_i v_{iR}/\sqrt 2$.
These trilinear terms also relate the origin of the $\mu$-term to the origin of neutrino masses and mixing angles, since neutrino Yukawa couplings 
$Y_{{\nu}_{ij}} \hat H_u\, \hat L_i \, \hat \nu^c_j$
 are present in the superpotential {generating Dirac masses for neutrinos, 
$m_{{\mathcal{D}_{ij}}}\equiv Y_{{\nu}_{ij}} {v_u}/{\sqrt 2}$.}
Remarkably, in the \mnSSM\ it is possible to accommodate neutrino masses
and mixings in agreement with experiments~\cite{Capozzi:2017ipn,deSalas:2017kay,deSalas:2018bym,Esteban:2018azc} via an EW seesaw
mechanism dynamically generated during the EW symmetry breaking~\cite{LopezFogliani:2005yw,Escudero:2008jg,Ghosh:2008yh,Bartl:2009an,Fidalgo:2009dm,Ghosh:2010zi,Liebler:2011tp}. The latter takes place 
through the couplings
$\kappa{_{ijk}} \hat \nu^c_i\hat \nu^c_j\hat \nu^c_k$,
giving rise to effective Majorana masses for RH neutrinos
${\mathcal M}_{ij}
= {2}\kappa_{ijk} \frac{v_{kR}}{\sqrt 2}$.
Actually, this is possible at tree level even with diagonal Yukawa couplings~\cite{Ghosh:2008yh,Fidalgo:2009dm}.
It is worth noticing here that the neutrino Yukawas discussed above also generate the effective bilinear terms $\mu_i \hat H_u\, \hat L_i $
with $\mu_i=Y_{{\nu}_{ij}} {v_{jR}}/{\sqrt 2}$,
used in the bilinear RPV model (BRPV)~\cite{Barbier:2004ez}.

We conclude therefore, that the \mnSSM\ solves not only the
$\mu$-problem, but also the $\nu$-problem, without
the need to introduce energy scales beyond the SUSY-breaking one.

The parameter space of the $\mn$, and in particular the
neutrino, neutral Higgs and stop sectors are 
relevant for our analysis in order to reproduce neutrino and Higgs data, and to obtain in the spectrum a stop as the LSP.
In particular, neutrino and Higgs sectors were discussed in Refs.~\cite{Kpatcha:2019gmq,Kpatcha:2019qsz,Kpatcha:2019pve,Heinemeyer:2021opc}, and we refer the reader to those works for details, although we will summarize the results below. First, we discuss here several simplifications that are convenient to take into account given the large number of parameters of the model.
{Using diagonal mass matrices for the scalar fermions, in order to avoid the
strong upper bounds upon the intergenerational scalar mixing (see e.g. Ref.~\cite{Gabbiani:1996hi}), from the eight minimization conditions with respect to $v_d$, $v_u$,
$v_{iR}$ and $v_{i}$ to facilitate the computation we prefer to eliminate
the
soft masses $m_{H_{d}}^{2}$, $m_{H_{u}}^{2}$,  
$m_{\widetilde{\nu}_{iR}}^2$ and
$m_{\widetilde{L}_{iL}}^2$
in favor
of the VEVs.
Also, we assume} for simplicity in what follows the flavour-independent couplings and VEVs $\lambda_i = \lambda$,
$\kappa_{ijk}=\kappa \delta_{ij}\delta_{jk}$, and $v_{iR}= v_{R}$. Then, the higgsino mass parameter $\mu$, bilinear couplings $\mu_i$ and Dirac and Majorana masses discussed above are given by:
\bea
\mu=3\la \frac{v_{R}}{\sqrt 2}, \;\;\;\;
\mu_i=Y_{{\nu}_{i}}  \frac{v_{R}}{\sqrt 2}, \;\;\;\;
m_{{\mathcal{D}_i}}= Y_{{\nu}_{i}} 
\frac{v_u}{\sqrt 2}, \;\;\;\;
{\mathcal M}
={2}\kappa \frac{v_{R}}{\sqrt 2},
\label{mu2}    
\eea
where 
we have already used the possibility of having diagonal neutrino Yukawa couplings $Y_{{\nu}_{ij}}=Y_{{\nu}_{i}}\delta_{ij}$ in the $\mn$ in order to reproduce neutrino physics.

\subsection{The neutrino sector}

\label{neutrino}

For light neutrinos, under the above assumptions, one can obtain
the following simplified formula for the effective mass matrix~\cite{Fidalgo:2009dm}:
\begin{eqnarray}
\label{Limit no mixing Higgsinos gauginos}
(m_{\nu})_{ij} 
\approx
\frac{m_{{\mathcal{D}_i}} m_{{\mathcal{D}_j}} }
{3{\mathcal{M}}}
                   \left(1-3 \delta_{ij}\right)
                   -\frac{v_{i}v_{j}}
                   {4M}, \;\;\;\;\;\;\;\;
        \frac{1}{M} \equiv \frac{g'^2}{M_1} + \frac{g^2}{M_2},         
\label{neutrinoph2}
  \end{eqnarray}     
where $g'$, $g$ are the EW gauge couplings, and $M_1$, $M_2$ the bino and wino soft {SUSY-breaking masses}, respectively.
This expression arises from the generalized EW seesaw of the $\mn$, where due to RPV the neutral fermions have the flavor composition
$(\nu_{i},\widetilde B^0,\widetilde W^0,\widetilde H_{d}^0,\widetilde H_{u}^0,\nu_{iR})$.
The first two terms in Eq.~(\ref{neutrinoph2})
are generated through the mixing 
of $\nu_i$ with 
$\nu_{iR}$-Higgsinos, and the third one 
also include the mixing with the gauginos.
These are the so-called $\nu_{R}$-Higgsino seesaw and gaugino seesaw, respectively~\cite{Fidalgo:2009dm}.
One can see from this equation that {once ${\mathcal M}$ is fixed, as will be done in the parameter analysis of Sec.~\ref{sec:parameter},
the most crucial independent parameters determining {neutrino physics} are}:
\bea
Y_{\nu_i}, \, v_{i}, \, M_1, \, M_2.
\label{freeparameters}
\eea
Note that this EW scale seesaw implies $Y_{\nu_i}\lsim 10^{-6}$
driving $v_i$ to small values because of the proportional contributions to
$Y_{\nu_i}$ appearing in their minimization equations. {A rough} estimation gives
$v_i\lsim m_{{\mathcal{D}_i}}\lsim 10^{-4}$.

{Considering the normal ordering for the neutrino mass spectrum,
and taking advantage of the 
dominance of the gaugino seesaw for some of the three neutrino families, three
representative type of solutions for neutrino physics using diagonal neutrino Yukawas were obtained in 
Ref.~\cite{Kpatcha:2019gmq}.
In our analysis we will use the so-called type 2 solutions, which have the structure
\bea
M>0, \, \text{with}\,  Y_{\nu_3} < Y_{\nu_1} < Y_{\nu_2}, \, \text{and} \, v_1<v_2\sim v_3,
\label{neutrinomassess}
\eea
In this case of type 2, it is easy to find solutions with the gaugino seesaw as the dominant one for the third family. Then, $v_3$ determines the corresponding neutrino mass and $Y_{\nu_3}$ can be small.
On the other hand, the normal ordering for neutrinos determines that the first family dominates the lightest mass eigenstate implying that $Y_{\nu_{1}}< Y_{\nu_{2}}$ and $v_1 < v_2,v_3$, {with both $\nu_{R}$-Higgsino and gaugino seesaws contributing significantly to the masses of the first and second family}. Taking also into account that the composition of the second and third families in the second mass eigenstate is similar, we expect $v_2 \sim v_3$. 
In Fig.~\ref{S1-NuParams-vs-Yvi} from Ref.~\cite{Kpatcha:2019gmq}, we show  
$\delta m^2=m^2_2-m^2_1$ 
versus $Y_{\nu_{i}}$ and $v_i$ for one of the scans carried out in that work, using
the results for normal ordering from Ref.~\cite{Esteban:2018azc}. 
As we can see, one obtains the hierarchy qualitatively discussed above for Yukawas and VEVs. 
We will carry out a similar analysis in Sec.~\ref{sec:results}, to correctly reproduce neutrino physics.

We will also argue in Sec.~\ref{sec:results} that the other two type of solutions of normal ordering for neutrino physics are not going to modify our results.
{The same conclusion is obtained in the case of working with the inverted 
ordering for the neutrino mass spectrum. The structure of the solutions is more involved for this case, because the two heaviest eigenstates are close in mass and the lightest of them has a dominant contribution from the first family. Thus, to choose $Y_{\nu_1}$ as the largest of the neutrino Yukawas helps to satisfy these relations. For the second and third family, a delicate balance between the contributions of $\nu_{R}$-Higgsino and gaugino seesaws is needed in order to obtain the correct mixing angles. In particular, a representative type of solutions for the case of inverted ordering has the structure $M>0$, with
$Y_{\nu_3} \sim Y_{\nu_2} < Y_{\nu_1}$, and $v_1<v_2\sim v_3$.}

\begin{figure}[t!]
  \centering
\includegraphics[width=\linewidth, height=0.4\textheight]{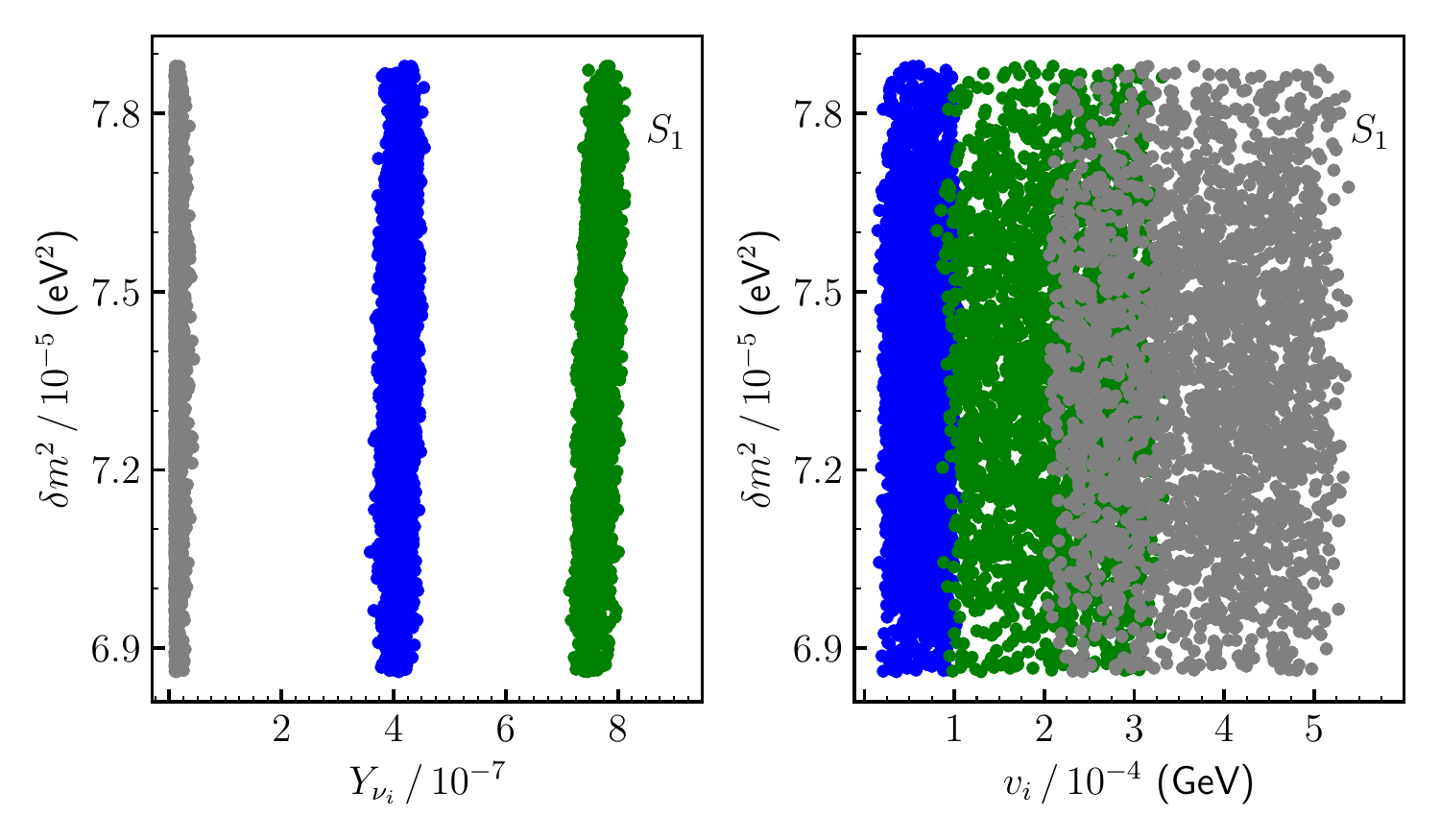}
\caption{$\delta m^2=m^2_2-m^2_1$ versus (left panel) neutrino Yukawas and (right panel) left sneutrino VEVs.
{Colors blue, green and grey correspond to $i=1,2,3$},
 respectively.}
 \label{S1-NuParams-vs-Yvi}
\end{figure}

\subsection{The Higgs sector}

The neutral Higgses are mixed with right and left sneutrinos, since
the neutral scalars and pseudoscalars in the $\mn$ have the flavor composition
$(H_{d}^0, H_{u}^0, \widetilde\nu_{iR}, \widetilde\nu_{i}) $.
Nevertheless, the left sneutrinos are basically decoupled from the other states, since
the off-diagonal terms of the mass matrix are suppressed by the small $Y_{\nu_{ij}}$ and $v_{i}$.
Unlike the latter states, the other neutral scalars can be substantially mixed.
Neglecting this mixing between 
the doublet-like Higgses and the three right sneutrinos, the expression of the tree-level mass of the { SM-like Higgs} is~\cite{Escudero:2008jg}:
\begin{eqnarray}
m_h^2 \approx 
m^2_Z \left(\cos^2 2\beta + 10.9\
{\lambda}^2 \sin^2 2\beta\right),
\end{eqnarray}
where $\tan\beta= v_u/v_d$, and $m_Z$ denotes the mass of the $Z$~boson.
Effects lowering (raising) this mass appear when the SM-like Higgs mixes with heavier (lighter) right sneutrinos. The one-loop corrections are basically determined by 
the third-generation soft {SUSY-breaking} parameters $m_{\widetilde u_{3R}}$, $m_{\widetilde Q_{3L}}$ and $T_{u_3}$,
where we have assumed for simplicity that for all soft trilinear parameters
$T_{ij}=T_{i}\delta_{ij}$.
These three parameters together with the coupling $\lambda$ and $\tan\beta$, are the crucial ones for Higgs physics. 
Their values can ensure that the model contains a scalar boson with a mass around $\sim 125 \gev$ and properties similar to the ones of the SM Higgs boson~\cite{Biekotter:2017xmf,Biekotter:2019gtq,Kpatcha:2019qsz,Biekotter:2020ehh}.

In addition, $\ka$, $v_R$ and the trilinear parameter $T_{\kappa}$ in the 
soft Lagrangian~(\ref{2:Vsoft}),
are the key ingredients to determine
the mass scale of the {right sneutrinos}~\cite{Escudero:2008jg,Ghosh:2008yh}.
For example, for $\lambda\lsim 0.01$ they are basically free from any doublet admixture, and using their minimization equations in the scalar potential
the scalar and pseudoscalar masses can be approximated respectively by~\cite{Ghosh:2014ida,Ghosh:2017yeh}:
\bea
m^2_{\widetilde{\nu}^{\mathcal{R}}_{iR}} \approx   \frac{v_R}{\sqrt 2}
\left(T_{\kappa} + \frac{v_R}{\sqrt 2}\ 4\kappa^2 \right), \quad
m^2_{\widetilde{\nu}^{\mathcal{I}}_{iR}}\approx  - \frac{v_R}{\sqrt 2}\ 3T_{\kappa}.
\label{sps-approx2}
\eea

Finally, $\lambda$ and the trilinear parameter $T_{\lambda}$ 
not only contribute to these masses
for larger values of $\lambda$, but
also control the mixing between the singlet and the doublet states and hence, they contribute in determining their mass scales as discussed in detail in Ref.~\cite{Kpatcha:2019qsz}.
We conclude that the relevant parameters
in the {Higgs (-right sneutrino) sector} are:
\bea
\lambda, \, \kappa, \, \tan\beta, \, v_R, \, T_\kappa, \, T_\lambda, \, T_{u_3}, \,  m_{\widetilde u_{3R}},
\, m_{\widetilde Q_{3L}}.
\label{freeparameterss}
\eea
Note
that the most crucial parameters for the neutrino sector~(\ref{freeparameters}) are basically decoupled from these parameters controlling Higgs physics.
This will become relevant in order to relax our demanding computing task, as will be discussed in 
Sec.~\ref{sec:parameter}
when analyzing the parameter space of the model.

\subsection{The stop sector}
\label{sec:stops0}

Similar to other squarks in the $\mn$, the mass matrix 
of the stops includes new terms with respect to the MSSM~\cite{Escudero:2008jg,Ghosh:2017yeh}. However, they are proportional to the small parameters $Y_{\nu_{ij}}$ and $v_{i}$, and therefore negligible. Thus, in practice the stop eigenstates of the $\mu\nu$SSM coincide basically with those of the MSSM.
In this approximation,
one obtains the following tree-level mass matrix in the flavor basis ($\widetilde t_L$, $\widetilde t_R$):
\begin{align}
m_{\widetilde{t}}^2= 
  \left( \begin{array}{cc} m^2_t +
m^2_{\widetilde{Q}_{3L}} + \Delta \widetilde u_L  & m_t X_t\\
       m_t X_t
 &  m^2_t + m^2_{\widetilde{u}_{3R}} +\Delta \widetilde u_R
         \end{array} \right),
\label{matrixsq2} 
\end{align}
where $m_t$ is the top-quark mass,
 $\Delta \widetilde u_{L,R}$ 
denote the D-term contributions
\bea
\Delta \widetilde u_L = m^2_Z \left( \frac{1}{2} - \frac{2}{3} \sin^2\theta_W \right) \cos 2\beta, \quad
\Delta \widetilde u_R =\frac{2}{3}  m^2_Z \sin^2\theta_W \cos 2\beta,
\label{delta-stop}
\eea
with $\theta_W$ the weak-mixing angle, and
$X_t$ the left-right stop mixing term
\bea
X_t = \left(\frac{T_{u_3}}{Y_{u_3}}-\frac{\mu}{\tan\beta}\right).
\eea

As can easily be deduced from Eq.~(\ref{matrixsq2}), the physical stops masses are controlled mainly by the value of the soft SUSY-breaking parameters:
\bea
m_{\widetilde Q_{3L}}, \, m_{\widetilde u_{3R}}, \, T_{u_3}.
\label{freeparametersstop}
\eea
Playing with their values, it is straightforward to obtain the lightest eigenvalue dominated either by the left stop composition ($\widetilde t_L$) or by the 
right stop composition ($\widetilde t_R$). This arises for small soft mass parameters or large left-right mixing. Note that in the case of the lightest stop mainly $\widetilde t_L$, a small value of the common soft mass $m_{\widetilde Q_{3L}}$ makes also $\widetilde b_L$ light although slightly heavier than 
$\widetilde t_L$ at tree level due to the D-term contribution,
$m_{\widetilde b_L}^{2} = m_{\widetilde t_{L}}^2
-m_W^2 \cos 2\beta$.
Nevertheless, depending on the value of $m_{\widetilde u_{3R}}$ and the left-right stop mixing controlled by $T_{u_3}$, the mass splitting between stop and sbottom can be much higher.}

\bigskip

\noindent
Note that the parameters in Eq.~(\ref{freeparametersstop}) are also crucial for obtaining the correct SM-like Higgs 
mass (see Eq.~(\ref{freeparameterss})). This will simplify our analysis in
Sec.~\ref{sec:results}.
There, 
we will sample the relevant parameter space of the $\mn$, which contains the independent parameters determining neutrino and
Higgs physics in Eqs.~(\ref{freeparameters}) and (\ref{freeparameterss}).
{Nevertheless, the parameters for neutrino physics $Y_{\nu_i}$, $v_{i}$, $M_1$ and $M_2$ are essentially decoupled from the parameters 
controlling Higgs physics.
Thus, for a suitable choice of the former parameters
reproducing neutrino physics, there is still enough freedom to reproduce in addition Higgs data by playing with 
$\lambda$, $\kappa$, $v_R$, $\tan\beta$, etc., 
as shown in Refs.~\cite{Kpatcha:2019gmq,Kpatcha:2019pve,Heinemeyer:2021opc}. 
As a consequence, we will not need to scan over most of the latter parameters, relaxing 
our demanding computing task.
Given the still large number of independent parameters} we have employed the 
{\tt Multinest}~\cite{Feroz:2008xx} algorithm as optimizer. {To compute
the spectrum and the observables we have used {\tt SARAH}~\cite{Staub:2013tta} to generate a 
{\tt SPheno}~\cite{Porod:2003um,Porod:2011nf} version for the model. }

\section{Stop LSP phenomenology}
\label{sec:stops}

The production of stops at colliders is dominated by QCD processes, since the RPV contributions to their production are strongly suppressed in the $\mn$. The pair production of colored SUSY particles at large hadron colliders has been extensively studied. Since we do not expect a significant difference from the values predicted in the MSSM, we make use of NNLL-fast-3.0~\cite{Beenakker:2016lwe,Beenakker:1997ut,Beenakker:2010nq,Beenakker:2016gmf} to calculate the number of stop pair events produced. In particular,
for our range of interest of stop masses between about 100 GeV and 2000 GeV, the production cross section is in the range between 1.42$\times 10^{3}$ pb 
and
 2$\times 10^{-5}$ pb.

\subsection{Decay modes}
\label{decays}

Concerning the stop decays, even though the stop eigenstates coincide basically with those of the MSSM scenario, novel differences appear in the decay cascades in the presence of RPV and additional states.
The interactions relevant for our analysis of the detectable decay of a stop LSP into SM particles, are given in Appendix~\ref{appendix}.
There, one can identify the most important contributions for the decays. 
In particular, the relevant diagram shown in Fig.~\ref{fig:stop_decay1}a corresponds to the term multiplying the projector $P_{L}$ and the
second term multiplying the projector $P_{R}$ in Eq.~(\ref{leptons}). {The diagram in
Fig.~\ref{fig:stop_decay1}b corresponds to the first term multiplying the projector $P_{L}$ and the
second and third terms multiplying the projector $P_{R}$ in Eq.~(\ref{neutrinos}).}
Thus the diagram in Fig.~\ref{fig:stop_decay1}a occurs mainly through the Yukawa couplings $Y_{t,b}$ 
of $\widetilde t$
with $b$ and charged higgsinos,
via the mixing between the latter and $\ell$.
The diagram in Fig.~\ref{fig:stop_decay1}b occurs through 
the 
gauge couplings $g$ and $g'$ of $\widetilde t$
with $t$ and 
neutral winos and binos, via the mixing between the latter and $\nu$.

\begin{figure}[t!]
 \centering
a)\includegraphics[scale=1,keepaspectratio=true]{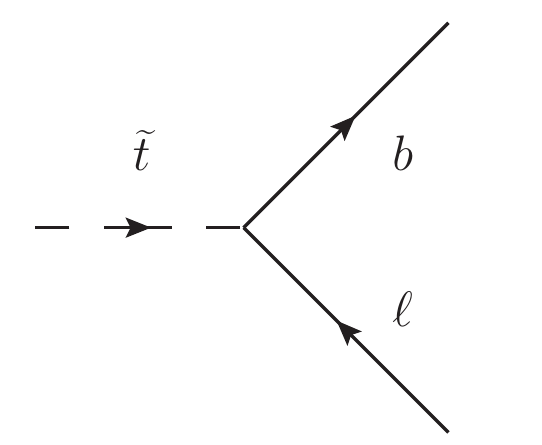}b)\includegraphics[scale=1,keepaspectratio=true]{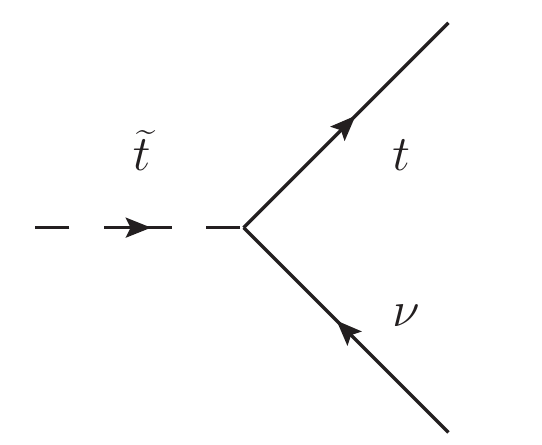}
 \caption{Dominant decay channels in the $\mn$ for a stop LSP. (a) bottom and leptons; (b) top and neutrinos. 
 }
 \label{fig:stop_decay1}
\end{figure}




In the case of {(pure)} left stop LSP or right stop LSP, the values of the partial decay widths can be approximated respectively,
as: 
\begin{eqnarray}
 \Gamma ({\widetilde t_L \to b\ell_i})\sim \frac{m_{\widetilde t}}{16\pi} \left(Y_b 
 \frac{\mu_i}
 {
 \mu}\right)^2,
 \quad  \sum_i \Gamma ({\widetilde t_L \to t\nu_i })\sim \frac{(m^2_{\widetilde t}-m^2_{t})^2 }{16\pi m^3_{\widetilde t}}\sum_i 
\left(\frac{g' }{6}U^V_{i4} +\frac{g}{2}U^V_{i5} \right)^2.
 \label{stL-decay}
\end{eqnarray}
\begin{eqnarray}
 \Gamma ({\widetilde t_R \to b \ell_i})\sim \frac{m_{\widetilde t}}{16\pi} \left(Y_t \frac{Y_{e_{i}}v_i }{
 \sqrt 2 \mu}\right)^2,
\quad \sum_i \Gamma ({\widetilde t_R \to t \nu_i })\sim \frac{(m^2_{\widetilde t}-m^2_{t})^2 }{16\pi m^3_{\widetilde t}}\sum_i 
 \left( \frac{2g'}{3} U^V_{i4} 
 \right)^2,
 \label{stR-decay}
 \end{eqnarray}
{where in the left formula of 
  Eq.~(\ref{stR-decay}) 
 the summation convention on repeated indices does not apply.}
 As discussed in Appendix~\ref{appendix}, $U^V$ is the matrix which diagonalizes the mass matrix 
 for the 
neutral fermions, and the above entries can 
be approximated as 
\begin{eqnarray}
U^V_{i4}\approx\frac{{-}g'}{M_1}\sum_{l}{\frac{v_{l}}{\sqrt{2}}U^{\text{\tiny{PMNS}}}_{il}},\quad \quad
U^V_{i5}\approx\frac{g}{M_2}\sum_{l}{\frac{v_{l}}{\sqrt 2}U^{\text{\tiny{PMNS}}}_{il}},
\label{--sneutrino-decay-width-2nus2}
\end{eqnarray}
where $U^{\text{\tiny{PMNS}}}_{il}$ are the entries of the PMNS matrix, with
$i$ and $l$ neutrino physical and flavor indices, respectively. 
 {We also approximate other entries of the matrices involved in the computation 
 (see Appendix~\ref{appendix}), as follows: $Z_{1,6}^{U}\approx 1$ for right stops and  $Z_{1,3}^{U}\approx 1$ for left stops, $U_{L, 33}^{d} \approx 1$ and $U_{R, 33}^{d} \approx 1$ (pure LH and RH bottom quarks), 
 $U_{L, 33}^{u} \approx 1$ and $U_{R, 33}^{u} \approx 1$ (pure LH and RH top quarks), $U_{L, j 4}^{e} \approx 0$ (vanishing charged wino-lepton mixing), $U_{i 7}^{V}\approx 0$ (vanishing higgsino $\widetilde H^0_u$-neutrino mixing),
 $U_{L, j 5}^{e}\approx {Y_{e_{i}}v_i }/{ \sqrt 2 \mu}$,
  $U_{R, j 5}^{e}\approx {\mu_i}/{\mu}$.}
 In addition we take $(m_{\tilde{t}}^{2}-m_{b}^{2})^2 \approx m_{\tilde{t}}^{4}$.

Depending on the composition of the stop, as well as the masses of higgsinos and gauginos, one decay channel or the other will dominate.
Let us remark nevertheless that the results of Sec.~\ref{sec:results}
have been obtained using the full numerical computation of decay widths,
taking also into account the contamination between left and right stops 
$|Z^U_{36}|^2$.
The latter turns out to be between 0.05\% and 3.6\% (0.07\% and 4\%) for the left stop (right stop) LSP points analyzed.

\subsection{LHC searches}
\label{sec:stops2}

There exist diverse searches at the LHC that can be applicable to constrain the event topologies described above. As shown in Fig.~\ref{fig:stop_decay1}, there are two possible situations: 
\begin{itemize}
\item The stop LSP decays producing a lepton ($e$, $\mu$ or $\tau$) and a bottom quark, giving rise to the process at the LHC shown in Fig~\ref{stop_decay_diagrams}a.
\item The stop LSP decays producing a neutrino and a top quark, giving rise to the process at the LHC shown in Fig~\ref{stop_decay_diagrams}b.
\end{itemize}
In both situations, with the proper lifetime spanning over a wide range. 
According to the search strategy to apply, we can further classify the signals in the following three cases,
applying for each of them different LHC searches depending on the lifetime scale:

\bigskip

\noindent
{\bf Case i) {Large ionization energy loss}}

\vspace{0.2cm}

\noindent 
Whenever the stop LSP is long-lived enough to measure ionization energy loss and/or time of flight, the searches for heavy charged long-lived particles are able to put strong constraints 
which are independent of the decay of the stop LSP. In particular, for each point of the parameter space with a stop LSP with proper lifetime greater than 3 m, the ATLAS search for heavy charged long-lived particles excludes stop masses below 1340 GeV~\cite{ATLAS:2019gqq}. 

\begin{figure}[t!]
\centering
a)\includegraphics[width=0.5\linewidth]{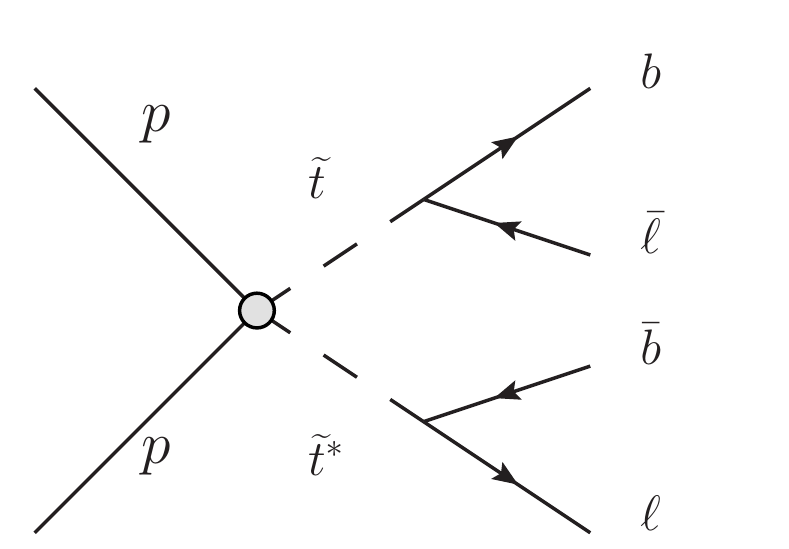}b)\includegraphics[width=0.5\linewidth]{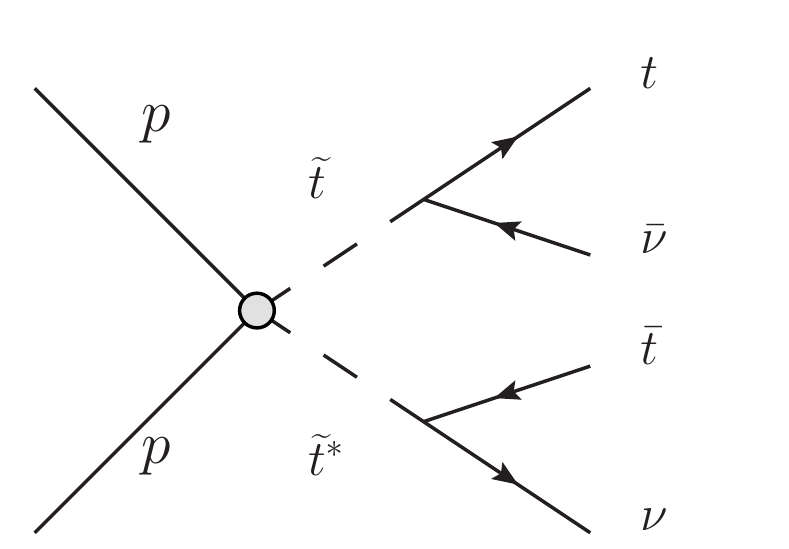}
  \caption{Processes relevant for the detection of the stop as the LSP at the LHC. }
\label{stop_decay_diagrams}
\end{figure}

\bigskip

\noindent
{\bf Case ii) {Displaced vertices}}

\vspace{0.2cm}

\noindent 
For shorter lifetime,
one can confront the points of the model with the limits from events with displaced vertices including leptons and/or jets. Let us consider first the situation with the decay of the stop LSP producing a lepton and a bottom quark at a significant distance of the first interaction point. In that case, one can compare the signal with the CMS searches~\cite{CMS:2017poa,CMS:2020iwv,CMS:2021kdm} which looks for pair-produced long-lived particles decaying to two $b$ quarks and two leptons. For each of the candidate points, one can compare the upper limit on the signal cross section from the plane $c\tau$-$m_{\widetilde{t}}$ with the cross section for the direct pair production of stops times the branching ratio (BR) of both stops decaying to leptons ($e$, $\mu$ or $\tau$) and bottom quarks. To complement the limits based on CMS results, we also use the limits obtained when confronting our data with the ATLAS search for heavy long-lived particles decaying to a bottom quark and a muon~\cite{ATLAS:2020xyo}.

The second situation corresponds to the decay of the stop LSP to a neutrino and a top quark. This signal can be constrained with searches for jets emerging from displaced vertices and missing transverse momentum, such as the search from ATLAS~\cite{ATLAS:2017tny}. This analysis, however, looks for a different signal produced in the decay of long-lived gluinos to two quarks and a neutral fermion through and off-shell heavier squark. To impose the limits on the stop LSP scenario, one can use the recasted version of the ATLAS search carried out in the work~\cite{Proceedings:2018het} and calculate the selection efficiency $\epsilon$ for events containing a pair of stops decaying to a top quark and a neutrino. Since the ATLAS search yields 0 background events in the SR, we impose an upper limit of 3 signal events, calculated as $\sigma(pp\to\widetilde{t}\widetilde{t}^*)\times BR(\widetilde{t}\to t\nu)^2\times \epsilon $.


\bigskip

\noindent
{\bf Case iii) {Prompt decays}}

\vspace{0.2cm}

\noindent 
If the proper lifetime of the stop LSP is too short to produce distinctive signals of long-lived particles, 
it can be constrained with the limits from searches for prompt signals.
In the first place, the prompt stop decay to leptons and bottom quarks can be constrained with the searches at the LHC for scalar leptoquarks. In particular, one can confront the stop LSP with the CMS search for pair production of first/second-generation scalar leptoquarks~\cite{CMS:2018ncu,CMS:2018lab}. For each of the points tested, one can compare $\sigma(pp\to\tilde{t} \tilde{t}^*)\times BR(\widetilde{t}\to b\ e/\mu)^2 $ with the upper limit on the $\tilde{t}\tilde{t}\to eebb/\mu\mu bb$ signal. In addition, one can apply the limit obtained from the ATLAS search for third-generation scalar leptoquarks~\cite{ATLAS:2021jyv} comparing $\sigma(pp\to\tilde{t} \tilde{t}^*)\times BR(\widetilde{t}\to b\tau)^2 $ with the upper limit on the $\tilde{t}\tilde{t}\to \tau\tau bb$ signal.

Secondly, the signal of the prompt decay of the stop LSP to a neutrino and a top quark is similar to the $R$-parity conserving (RPC) decay of the stop producing a neutralino LSP in the context of the MSSM, in the limit of massless neutralino. Thus one can constrain the points where the decay is dominated by the $t\nu$ channel with the ATLAS search for stops in final state with jets plus missing transverse momentum~\cite{ATLAS:2017drc,ATLAS:2020dsf}.
For each of the points analyzed, one can compare the upper limit calculated by ATLAS with $\sigma(pp\to\widetilde{t}\widetilde{t}^*)\times BR(\widetilde{t}\to t\nu)^2 $. This constraint can be complemented with the restriction obtained similarly from the corresponding CMS search for stops~\cite{CMS:2021eha}.

{Finally, let us comment that in the ATLAS search for third-generation scalar leptoquarks~\cite{ATLAS:2021jyv}, the combination of the decays to $b\tau$ and to $t\nu$ is considered for different values of the corresponding BRs. However, the limits obtained over the  mass are only mildly dependent on the BRs.}

\medskip

\noindent

\section{Strategy for the scanning}
\label{strategy}

In this section, we describe the methodology that we have employed to search for points of our
parameter space that are compatible with the current experimental data on neutrino and Higgs physics, as well as ensuring that the stop is the LSP.
In addition, we have demanded the compatibility with some flavor observables,
such as $B$ and $\mu$ decays.
To this end, we have performed scans on the parameter space of the model, with the input parameters optimally chosen.
It is worth noting here that we will not make any statistical interpretation of the set of points obtained, i.e. the {\tt Multinest} algorithm is just used to obtain viable points.

\subsection{Experimental constraints}
\label{sec:constr}

All experimental constraints (except the LHC searches which are discussed in the previous section) are taken into account as follows:

\begin{itemize}

\item Neutrino observables\\
We have imposed the results for normal ordering from Ref.~\cite{Esteban:2018azc}, selecting points from the scan that lie within $\pm 3 \sigma$ of all neutrino observables. On the viable obtained points we have imposed the cosmological upper
bound on the sum of the masses of the light active neutrinos given
by $\sum m_{\nu_i} < 0.12$ eV~\cite{Aghanim:2018eyx}.

\item Higgs observables\\
The Higgs sector of the $\mn$ is extended with respect to the (N)MSSM.
For constraining the predictions in that sector of the model, we have interfaced 
{\tt HiggsBounds} {{v}}5.3.2~{\cite{Bechtle:2008jh,Bechtle:2011sb,Bechtle:2013wla,Bechtle:2015pma,Bechtle:2020pkv}} with {\tt Multinest}, using a 
conservative $\pm 3 \gev$ theoretical uncertainty on the SM-like Higgs boson in the $\mn$ as obtained with {\tt SPheno}. 
Also, in order to address whether a given Higgs scalar of the $\mn$ 
is in agreement with the signal observed by ATLAS and CMS, we have interfaced
{\tt HiggsSignals} {{v}}2.2.3~{\cite{Bechtle:2013xfa,Bechtle:2020uwn}} with {\tt Multinest}.
We require that the $p$-value reported by {\tt HiggsSignals} be larger than 5\%.

\item $B$~decays\\
$b \to s \gamma$ occurs in the SM at leading order through loop diagrams.
We have constrained the effects of new physics on the rate of this 
process using the average {experimental value of BR$(b \to s \gamma)$} $= (3.55 \pm 0.24) \times 10^{-4}$ provided in Ref.~\cite{Amhis:2012bh}. 
Similarly to the previous process, $B_s \to \mu^+\mu^-$ and  $B_d \to \mu^+\mu^-$ occur radiatively. We have used the combined results of LHCb and CMS~\cite{CMSandLHCbCollaborations:2013pla}, 
$ \text{BR} (B_s \to \mu^+ \mu^-) = (2.9 \pm 0.7) \times 10^{-9}$ and
$ \text{BR} (B_d \to \mu^+ \mu^-) = (3.6 \pm 1.6) \times 10^{-10}$. 
We put $\pm 3\sigma$ cuts from $b \to s \gamma$, $B_s \to \mu^+\mu^-$ and $B_d \to \mu^+\mu^-$, {as obtained with {\tt SPheno}}. {We have also checked that the values obtained are compatible with the $\pm 3 \sigma$ of the recent results from the LHCb collaboration \cite{Santimaria:2021}. 
}

\item $\mu \to e \gamma$ and $\mu \to e e e$\\
We have also included in our analysis the constraints {from 
BR$(\mu \to e\gamma) < 4.2\times 10^{-13}$~\cite{TheMEG:2016wtm}}
and BR$(\mu \to eee) < 1.0 \times 10^{-12}$~\cite{Bellgardt:1987du}, {as obtained with {\tt SPheno}}.


\item Chargino mass bound\\
{Charginos have been searched at LEP with the result of a lower limit on the lightest chargino mass of 103.5 GeV in RPC MSSM, assuming universal gaugino and sfermion masses at the GUT scale and electron sneutrino mass larger than 300 GeV~\cite{wg1}. This limit is affected if the mass difference between chargino and neutralino is small, and the lower bound turns out to be in this case 92 GeV~\cite{wg2}. LHC limits  can be stronger but for very specific mass relations~\cite{ATLAS:2019lff,CMS:2018xqw,ATLAS:2017qwn,CMS:2018yan}.
Although in the $\mn$ there is RPV and therefore these constraints do not apply automatically, we have chosen in our analysis a conservative limit of $m_{\widetilde \chi^\pm_1} > 92 \gev$.}

\end{itemize}




\subsection{Parameter analysis}
\label{sec:parameter}

In order to relax our demanding computing task, we will perform scans using the fewest possible parameters to get points with the stop as LSP satisfying the experimental constraints discussed above.

First, the soft stop masses $m_{\widetilde Q_{3L}}$ and $m_{\widetilde u_{3R}}$, and the trilinear parameter $T_{u_3}$ are very important in our analysis, since they not only control the physical stop masses but are also crucial for obtaining the correct SM-like Higgs mass, as discussed in Sec.~\ref{sec:model}. In addition, reproducing Higgs data requires suitable additional parameters $\lambda$, $\kappa$, $\tan\beta$, $v_R$, $T_\kappa$ and $T_\lambda$
(see Eq.~(\ref{freeparameterss})). Nevertheless, we can fix $\kappa$, $T_\kappa$ and $v_R$, 
which basically control the right sneutrino sector, while suitably varying $\lambda$, $T_\lambda$, and $\tan\beta$. 
Concerning neutrino physics, as discussed also in Sec.~\ref{sec:model} the most crucial parameters~(\ref{freeparameters}) are basically decoupled from those controlling Higgs physics~(\ref{freeparameterss}).
Thus, for a suitable scan of $Y_{\nu_i}$, $v_{i}$, $M_1$ and $M_2$ reproducing neutrino physics, there is still enough freedom to reproduce in addition Higgs data by playing with concrete values of  $\lambda$, $\kappa$, $\tan\beta$, $v_R$, etc. 
In this regard, in order to optimize the parameters of the scan, we will only scan over the VEVs $v_1$ and $v_2$ and take the third left sneutrino VEV as $v_3 = 1.8\ v_2$, a relation that is inspired by the specific solution for neutrino physics
discussed in Eq.~(\ref{neutrinomassess})~\cite{Kpatcha:2019gmq}.
Also we will scan over $Y_{\nu_{1}}$, take $Y_{\nu_{2}} = 1.8\ Y_{\nu_{1}}$ and fix $Y_{\nu_{3}}$,
which allows us to reproduce the normal ordering of neutrino masses with $Y_{\nu_3} < Y_{\nu_1} <  Y_{\nu_2}$.
Finally, we will scan over the soft bino mass $M_1$, which is sufficient to reproduce neutrino data.

Summarizing, we will perform  two scans over the 9~parameters, corresponding to left stop as LSP and right stop as LSP, respectively, as shown in Table~\ref{Scans-parameters}.
For Scan 1, the soft mass of the squark doublet is taken as $m_{\widetilde Q_{3L}} \in (200, 1200)$ GeV, whereas the soft mass of the right stop is fixed to a large value, $m_{\widetilde u_{3R}} = 2000$ GeV, as shown in 
Table~\ref{Fixed-parameters}.
Scan 2 is similar but with the replacement $m_{\widetilde Q_{3L}}\leftrightarrow m_{\widetilde u_{3R}}$.
The lower bound 
of 200 GeV is chosen to avoid too light stops/sbottoms, and the upper bound of 1.2 TeV seems to be a reasonable value for the analysis of the stop as the LSP. 
We choose a range of moderate/large values of $\lambda \in (0.3, 07)$, thus we are in a similar situation as in the NMSSM and small/moderate values of $\tan\beta$, $|T_{u_{3}}|$, and soft stop masses, as shown in Table~\ref{Scans-parameters}, are necessary to obtain through loop effects the correct SM-like Higgs mass~{\cite{Biekotter:2017xmf,Biekotter:2019gtq,Kpatcha:2019qsz,Biekotter:2020ehh}}.
Given the range of soft stop masses considered, we take the upper bound of $-T_{u_{3}}$ at 2 TeV to be able to obtain in our moderate/large $\lambda$ limit {the appropriate mixing reproducing the SM-like Higgs mass.}
The parameter $T_{\lambda}$ is relevant to obtain the correct values of the off-diagonal terms of the mass matrix mixing the right sneutrinos with Higgses, and we choose an appropriate range for that task.
Concerning $\tan\beta$, we consider a range of possible values that are simultaneously compatible with Higgs physics and the stop as LSP.
The ranges of $v_{2,\,3}$ and $Y_{\nu_1}$ are natural in the context of the EW scale seesaw of the  $\mn$~\cite{Kpatcha:2019gmq,Kpatcha:2019pve,Heinemeyer:2021opc}. 
Finally, 
we choose the range of $M_1$ such that the lightest bino is heavier than the stop LSP.

\begin{table}
\begin{center}
\begin{tabular}{|l|l|}
\hline
 \multicolumn{1}{|c|}{\bf Scan 1 }&\multicolumn{1}{c|}{ \bf Scan 2}\\
\cline{1-2}
   $m_{\widetilde Q_{3L}} \in (200, 1200)$ &  $ m_{\widetilde u_{3R}} \in (200, 1200)$\\ 
\hline
\multicolumn{2}{|c|}{ $- T_{u_{3}} \in  (0, 2000)$ }\\
\multicolumn{2}{|c|}{ $Y_{\nu_{1}} \in (10^{-8} , 10^{-6})$ }\\
\multicolumn{2}{|c|}{ $v_{1,2} \in (10^{-5} , 10^{-3})$  }\\
\multicolumn{2}{|c|}{ $M_1 \in (1500 , 2500)$ }\\
\multicolumn{2}{|c|}{ $T_\lambda  \in  (0.5 , 2000)$ }\\
\multicolumn{2}{|c|}{ $\lambda \in (0.3 , 0.7)$ }  \\ 
\multicolumn{2}{|c|}{ $\tan\beta \in (1 , 20)$  }  \\ 
\cline{1-2}
\end{tabular}
\end{center}
  \caption{Range of low-energy values of the input parameters that are varied in the two scans. 
  The VEVs $v_i$ and the soft SUSY-breaking parameters $m_{\widetilde Q_{3L}}$, $ m_{\widetilde u_{3R}}$, $T_{u_{3}}, T_\lambda$ and $M_1$ are given in GeV. Relations $v_3 = 1.8 \, v_2$ and $Y_{\nu_{2}} = 1.8 \, Y_{\nu_{1}}$ are used.}
 \label{Scans-parameters}
\end{table} 

\begin{table}
\begin{center}
\begin{tabular}{|l|l|}
\hline
 \multicolumn{1}{|c|}{\bf Scan 1 }&\multicolumn{1}{c|}{ \bf Scan 2}\\
\cline{1-2}
   $m_{\widetilde u_{3R}} = 2000 $ &  $m_{\widetilde Q_{3L}}  = 2000$\\
\hline
\multicolumn{2}{|c|}{ $\kappa$ = 0.5} \\ 
\multicolumn{2}{|c|}{ $-T_{\kappa}$ = 300 }\\ 
\multicolumn{2}{|c|}{ $Y_{\nu_{3}}= 10^{-8}$ }\\ 
\multicolumn{2}{|c|}{ $\sqrt{2} v_R$ = 3500} \\ 
\multicolumn{2}{|c|}{  $M_2$ = 2000, $ M_3$ = 2700} \\ 
\multicolumn{2}{|c|}{$m_{\widetilde Q_{1,2\,L}}=m_{\widetilde u_{1,2\,R}}=m_{\widetilde e_{1,2,3\,R}}=$ 2000 } \\
\multicolumn{2}{|c|}{ $T_{u_{1,2}}=T_{d_{1,2}}=T_{e_{1,2}}=0$} \\
\multicolumn{2}{|c|}{$T_{d_{3}}=$ 100, $T_{e_{3}}=$ 40 }\\
\multicolumn{2}{|c|}{ $-T_{\nu_{1,2,3}}= 10^{-3}$}    \\ 
\cline{1-2}
\end{tabular}
\end{center}
  \caption{Low-energy values of the input parameters that are fixed in the two scans. The VEV $v_R$ and the soft trilinear parameters, soft wino and gluino masses and soft scalar masses are given in GeV.}
 \label{Fixed-parameters}
\end{table}

Other benchmark parameters relevant for Higgs physics are fixed to appropriate values, as shown in Table~\ref{Fixed-parameters}.
To ensure that chargino is heavier than stop, the lower value of $\lambda$ forces us to choose a large value for $v_R$ in order to obtain a large enough value of $\mu$ (see Eq.~(\ref{mu2})). Thus we choose $\sqrt{2} v_R=3500$ GeV implying a range for $\mu\in (1575, 3675)$ GeV. 
The parameters $\ka$ and $T_{\kappa}$ are crucial to determine the mass scale of the right sneutrinos.
We choose $T_{\kappa}=-300 \gev$ to have heavy pseudoscalar {RH} sneutrinos (of about 1255~GeV), and therefore 
the value of $\kappa$ has to be large enough in order to avoid 
too light (even tachyonic) scalar right sneutrinos. Choosing $\kappa=0.5$, we get masses for the latter of about $1593\gev$. {This value of $\kappa$ (and the one chosen above for $v_R$) also implies that the Majorana mass is fixed to ${\mathcal M}=1750$ GeV.
The values of the remaining parameters shown in Table~\ref{Fixed-parameters} concern slepton, squark and  gaugino masses, as well as quark and lepton trilinear parameters, which are not specially relevant for our analysis.
The values for $T_{\nu_{i}}$ are chosen to obtain left sneutrinos heavier than the stop LSP (see the discussion about their masses in Refs.~\cite{Kpatcha:2019gmq,Kpatcha:2019qsz,Kpatcha:2019pve,Heinemeyer:2021opc}). They are reasonable within the supergravity framework, where
the trilinear parameters are proportional to the corresponding Yukawa couplings, i.e. in this case $T_{\nu_{i}}= A_{\nu_{i}} Y_{\nu_{i}}$ implying $-A_{\nu_{i}}\in$ (555, $10^{5}$) GeV. 
In a similar way, the values of $T_{d_3}$ and $T_{e_3}$ have been chosen taking into account the corresponding Yukawa couplings.

{However, as we will see in the next section, the comparison of the results of the scans above with LHC data~\cite{CMS:2017poa,ATLAS:2017tny,Proceedings:2018het} 
discussed 
in Sec.~\ref{sec:stops2}, implies that all points found compatible with the experimental constraints of Sec.~\ref{sec:constr}, turn out to be excluded.
Given this strong constraint, we explored a bigger mass range for the stop
in order to obtain points allowed by LHC data.
To carry out this analysis we used the following strategy.
We combined the previous scans of the relevant parameters, with a subsequent intelligent search to increase the number of points obtained. For this purpose, we took advantage again that the parameters controlling neutrino and Higgs physics are basically decoupled. Thus, we chose first several starting points for the analysis with fixed values of the parameters allowing to fulfill neutrino physics, i.e. $Y_{\nu_{1}}$, $v_{1}$, $v_{2}$ and $M_1$, taking besides for these points $Y_{\nu_{2,3}}$, $v_3$, $M_2$ as in Tables~\ref{Scans-parameters} and~\ref{Fixed-parameters}.
Then, we scanned over a subset of the parameters of Table~\ref{Scans-parameters} relevant for Higgs physics and left (right) stop as the LSP, by varying $m_{\widetilde Q_{3L}}$ ($m_{\widetilde u_{3R}}$) and $m_{\widetilde u_{3R}}$ ($m_{\widetilde Q_{3L}}$) in a new range of $900-2000$ GeV, and $2000-3000$ GeV, respectively, and by extending the value of $T_{u_3}$ up to $-3000$ GeV.
For the rest of the parameters of Table~\ref{Scans-parameters}, i.e. $\lambda, T_\lambda$ and $\tan\beta$, we chose concrete appropriate values for each of the starting points of the analysis.
Additionally, in order to keep the stop as the LSP, we modified in Table~\ref{Fixed-parameters} the values of the parameters  
$T_\kappa$ and $T_{\nu}$. Thus, we used $T_\kappa = - 800$ GeV 
to increase the mass of pseudoscalar right sneutrinos, 
and 
$T_{\nu_{1,2,3}} = - 10^{-2}$ GeV to increase the masses of left sneutrinos and sleptons.}


\section{Results}
\label{sec:results}



Following the methods described in the previous sections, in order
to find regions consistent with experimental observations we performed scans, and our results are presented here.
To carry this analysis out, we selected first points from the scans that lie within $\pm 3\sigma$ of all neutrino physics observables \cite{Esteban:2018azc}.
Second, we put $\pm 3\sigma$ cuts from $b \to s \gamma$, $B_s \to \mu^+\mu^-$ and $B_d \to \mu^+\mu^-$ 
and require the points to satisfy also the upper limits of $\mu \to e \gamma$ and $\mu \to eee$. 
In the third step, we imposed that Higgs physics is realized.
In particular, we require that the p-value reported by {\tt HiggsSignals} be larger
than 5\%.
Also, since we are interested in the left stop as LSP or the right stop as LSP, of the allowed points we selected those satisfying these conditions. 

We show in Fig.~\ref{LHC-mt1vsctau2} the proper decay length of the stop LSP 
for the points of the parameter space fulfilling the above experimental constraints.
In the case of the right stop LSP, there are light-green points with
decay lengths between {6.8 m} and 3 m, and with the corresponding masses
between {100 GeV and 176 GeV}. Following the discussion of Case {\it (i)} in Sec.~\ref{sec:stops2}, all these points with such low masses (and therefore with large production cross sections)
are excluded.

\begin{figure}[t!]
\centering
\includegraphics[width=\linewidth, height= 0.35\textheight]{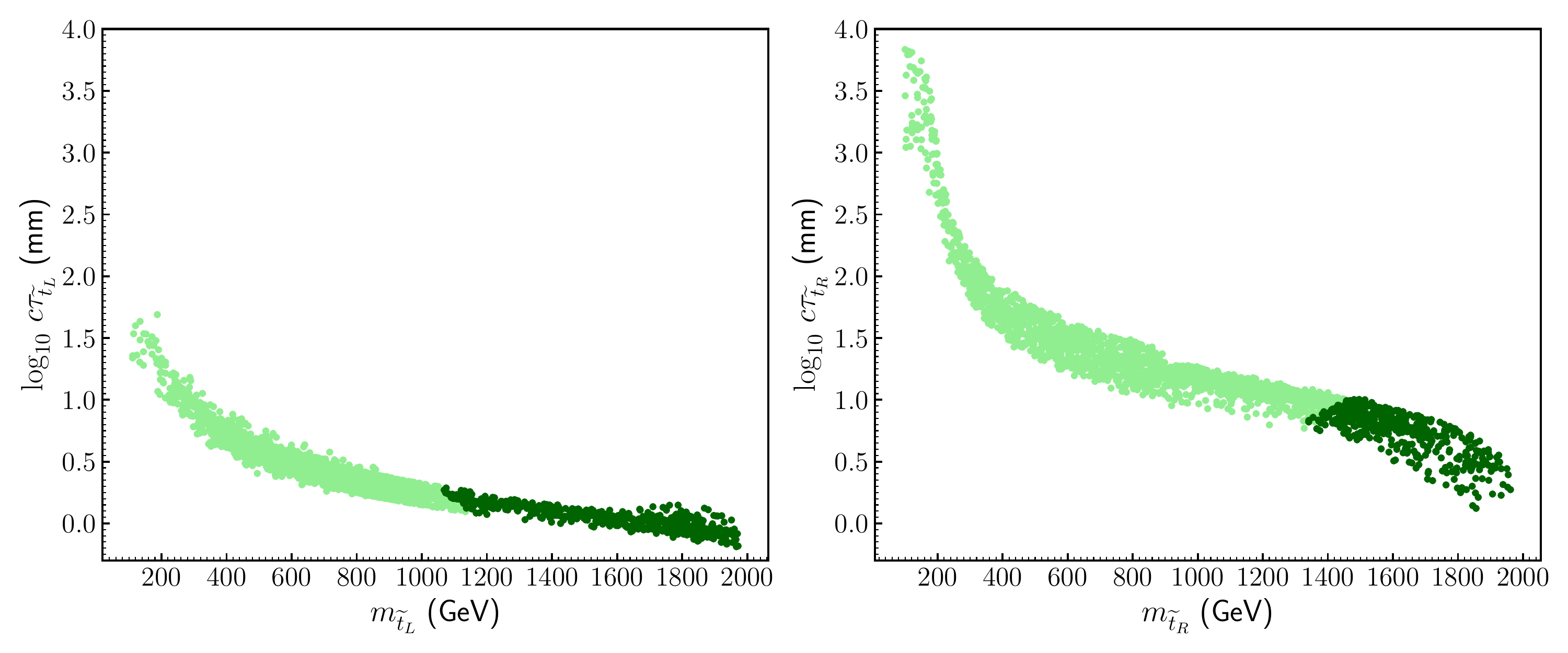}
  \caption{Proper decay length $c\tau_{\widetilde t}$ versus the stop mass $m_{\widetilde t}$ for (left plot)
  left stop LSP and (right plot) right stop LSP.
  All points fulfill the experimental constraints of Sec.~\ref{sec:constr}. 
  Light (dark)-green points do not fulfill (fulfill) the LHC constraints.}
\label{LHC-mt1vsctau2}
\end{figure}

\begin{figure}[t!]
\centering
\includegraphics[width=\linewidth, height= 0.35\textheight]{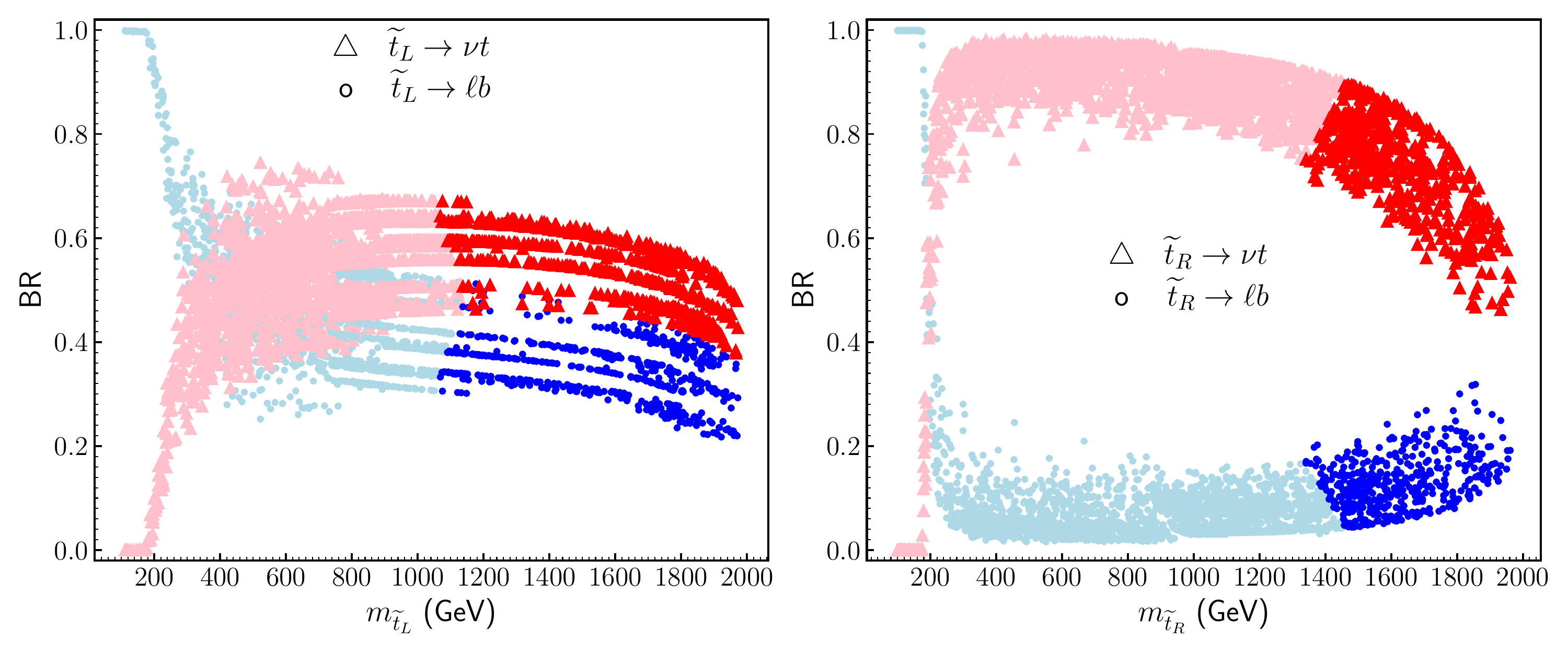}
  \caption{Points of Fig.~\ref{LHC-mt1vsctau2}, but showing for them the
  sum of branching ratios of stop LSP decaying to 
  $\ell b$ (blue points)
  and 
  $\nu t$ (red triangles) versus the stop mass $m_{\widetilde t}$.
  Light (dark)-red and blue points do not fulfill (fulfill) the LHC constraints.
  }
\label{LHC-mt1vsBRleptonb2}
\end{figure}

 For the rest of the points in our scans, the decay lengths are displaced and therefore we can apply for all of them the LHC constraints discussed in Case {\it (ii)} of Sec.~\ref{sec:stops2}. 
{In 
Fig.~\ref{LHC-mt1vsctau2}, for the left stop LSP the light-green points correspond to decay lengths between 48.9 mm and 1.24 mm with stop masses between 111 GeV and 1134 GeV.
For the right stop LSP, the light-green points have decay lengths as small as 5.9 mm and masses as large as 1450 GeV.
In Fig.~\ref{LHC-mt1vsBRleptonb2}, we show for these points the sum of BRs for the decay of a stop LSP into 
leptons ($e$, $\mu$ or $\tau$) and bottom quark with light-blue colors, and neutrinos and top quark with light-red colors. As we can see, for stop masses around the top mass the dominant decay channel corresponds to $\ell b$, as can also be deduced from Eqs.~(\ref{stL-decay}) 
and~(\ref{stR-decay}). However, for larger masses left and right stop behave in a different way. Whereas for the left stop the BRs are of similar order, for the right stop the BR to 
$\nu t$ is much larger than the one to $\ell b$.
The comparison of these results with LHC data~\cite{CMS:2017poa,ATLAS:2017tny,Proceedings:2018het} discussed 
in Sec.~\ref{sec:stops2}, implies that all points turn out to be excluded.}

{Dark-green points allowed by LHC data start to appear
for left (right) stop LSP masses of 1068 GeV (1341 GeV), 
corresponding to a decay length of 1.86 mm (6.61 mm) as shown in Fig.~\ref{LHC-mt1vsctau2}. The BRs for these points are also shown in 
Fig.~\ref{LHC-mt1vsBRleptonb2} with dark-red and dark-blue colors.
For low stop masses, the combination of the two decay channels discussed in Fig.~\ref{stop_decay_diagrams} takes account of almost the total decay amplitude of the stop. However, when the mass of the stop approaches the mass of the lightest higgsino, the decay $\widetilde{t}\to h t \bar{\nu}$, mediated by an off-shell higgsino, has a non negligible contribution. This effect can be observed in Fig.~\ref{LHC-mt1vsBRleptonb2} for masses larger than about 1600 GeV, where the shown branching fractions no longer combine to give 1.
No visible signal associated to this decay is possible, since the production cross section is too small.}  
On the other hand, as can be seen in Fig.~\ref{LHC-mt1vsctau2} the value of $c\tau$ decreases with increasing values of the stop mass. Therefore, for  sufficiently heavy masses, the stop becomes short lived. {However, this happens for stop masses above the lower limit of 1300 GeV (1500 GeV)} obtained by prompt stop~\cite{ATLAS:2017drc,ATLAS:2020dsf,CMS:2021eha} (scalar leptoquark~\cite{CMS:2018ncu,CMS:2018lab,ATLAS:2021jyv}) searches,
 discussed in Case {\it (iii)} of Sec.~\ref{sec:stops2}.
 Thus, we do not expect prompt searches to constrain the stop as LSP in the $\mn$.

{The HL-LHC is expected to be sensitive to heavier stop masses. For example, the search for promptly decaying top squarks in the $\tilde{t} \tilde{t} \to t t \tilde{\chi}^0_1 \tilde{\chi}^0_1$ channel can discover (exclude) stops up to masses of 1.25 (1.7)~TeV with 3~ab$^{-1}$ for the massless neutralino LSP~\cite{CidVidal:2018eel}. As seen in Fig.~\ref{LHC-mt1vsctau2}, for these masses the stops can still have a sizable decay length, and therefore it would be better to relax the requirement for prompt decay. In addition, the reconstruction of displaced decay vertices may be useful to improve the sensitivity, as in the case of metastable gluino searches~\cite{Ito:2017dpm, Ito:2018asa}. For the $\tilde{t} \to \ell b$ channel, on the other hand, the leptoquark searches will again be useful (with a relaxation of the prompt-decay requirement). For example, the discovery reach of the search for leptoquarks in the $\tau \bar{\tau} b \bar{b}$ final state at the HL-LHC is $\sim 1200$ (1500)~GeV with the 300~fb$^{-1}$ (3~ab$^{-1}$)~\cite{CidVidal:2018eel}. We, however, recall that the expected reach should be worse in our scenario as the $\tilde{t} \to \ell b$ channel is always subdominant, as seen in Fig.~\ref{LHC-mt1vsBRleptonb2}. Again, the reconstruction of displaced decay vertices may be able to improve the discovery reach significantly. It is worth noting that the detection of this decay channel strongly favors our model against RPC SUSY scenarios such as the MSSM. Of course, even when we discover the $\tilde{t} \to \ell b$ signature, we still need further investigation to distinguish our model from leptoquark models. We, however, point out that in usual leptoquark scenarios, there is no reason for the leptoquarks to be metastable (unless the relevant coupling is taken to be extremely small by hand), while in our model a sizable decay length of the top squark LSP is predicted. The reconstruction of displaced decay vertices, therefore, provides an useful way to discriminate our model against leptoquark models.  }


{Let us finally remark that 
the use of other type of solutions for neutrino physics different from the one presented in
Eq.~(\ref{neutrinomassess}),
would not modify the results obtained.
In particular, note that the CMS search~\cite{CMS:2017poa} shows similar sensitivities to the signals including the decay of the stop LSP to electrons, muons or taus. Therefore, if we move to regions of the parameter space with different solutions for neutrino physics, which would imply different relative sizes of the partial decay widths to leptons, the results will not be affected.
In addition, Fig.~\ref{LHC-mt1vsBRleptonb2} shows that the two terms of Eq.~(\ref{stL-decay}) are balanced. If a change of the type of neutrino solution results in one of the decay channels being more dominant, the corresponding process will be more excluded.}


\section {Conclusions}
\label{sec:conclusion}

In the framework of the RPV $\mn$ model, we analyzed the signals expected at the LHC for a stop as the LSP.
The current experimental data on neutrino and Higgs physics, as well as flavor observables such as 
$B$ and $\mu$ decays, were imposed on the parameter space of the model.
The stops are pair produced, decaying with displaced vertices to a lepton and a bottom quark or a neutrino and a top quark. 
In the case of the left (right) stop LSP, the decay lengths can be as large as 4.89 cm (680 cm)  for masses as small as 111 GeV (100 GeV).
We compared the predictions of this scenario with ATLAS and CMS searches for long-lived particles~\cite{ATLAS:2019gqq,CMS:2017poa,CMS:2020iwv,CMS:2021kdm,ATLAS:2020xyo,ATLAS:2017tny,Proceedings:2018het,CMS:2018ncu,CMS:2018lab,ATLAS:2021jyv,ATLAS:2017drc,ATLAS:2020dsf,CMS:2021eha}.
Our results translate into a lower limit on the mass of the left (right) stop LSP of {1068 GeV (1341 GeV)}, corresponding to a decay length of 
{1.86 mm (6.61 mm).}

We expect that more parameter points will be explored at the HL-LHC. The sensitivities of the stop searches may significantly be improved if the search strategies are optimized for long-lived particles; for this purpose, a dedicated trigger for long-lived particles may be useful~\cite{Alimena:2021mdu}. As we have discussed above, $\mathcal{O}(1)$~mm displaced vertices associated with the stop decay signatures ($\tilde{t} \to \nu t$ or $\tilde{t} \to \ell b$) are the smoking gun signals for the present scenario, and the discovery of such a signature will strongly favor our model against the RPC SUSY scenarios.


\begin{acknowledgments}

The work of E.K.and C.M. was partially supported by the Spanish Research Agency through the grants IFT Centro de Excelencia 
Severo Ochoa No CEX2020-001007-S and
PGC2018-095161-B-I00.
They are funded by MCIN/AEI/10.13039/501100011033 and the second grant also by ERDF "A way of making Europe".
The work of E.K. was supported by the grant "Margarita Salas" for the training of young doctors (CA1/RSUE/2021-00899), co-financed by the Ministry of Universities, the Recovery, Transformation and Resilience Plan, and the Autonomous University of Madrid.
The work of I.L.\ was funded by the Norwegian Financial Mechanism 2014-2021, grant DEC-2019/34/H/ST2/00707. 
The work of D.L.\ was supported by the Argentinian CONICET, and also 
through PIP 11220170100154CO. 
The work of N.N.\ was supported in part by the Grant-in-Aid for
Young Scientists (No.21K13916), Innovative Areas (No.18H05542), and Scientific Research B (No.20H01897).

\end{acknowledgments}


\appendix
\numberwithin{equation}{section}
\numberwithin{figure}{section}
\numberwithin{table}{section}

\numberwithin{equation}{subsection}
\numberwithin{figure}{subsection}
\numberwithin{table}{subsection}

\section{One Up Squark-two Fermion--Interactions}
\label{appendix}

In this Appendix we write the relevant interactions for our computation of the decays of the stop LSP, 
following {\tt SARAH} notation~\cite{Staub:2013tta}.
In particular, 
now
$a,b=1,2,3$ are family indexes, $i,j,k$ are the indexes for the physical states, and $\alpha,\beta,\gamma=1,2,3$ are $SU(3)_C$ indexes.
The matrices $Z^U, U^d_{L,R}, U^u_{L.R}, U^e_{L,R}$ and $U^V$ diagonalize the mass matrices of up squarks, down quarks, up quarks, charged fermions (leptons, gauginos and higgsinos) and neutral fermions (LH and RH neutrinos, gauginos and higgsinos), respectively.
More details about these matrices can be found in Appendix B of Ref.~\cite{Ghosh:2017yeh}.
Taking all this into account, in the basis of 4--component spinors with the projectors 
$P_{L,R}=(1\mp\gamma_5)/2$, the interactions for the mass eigenstates are as follows.


\subsection{\bf Up squark-down quark-lepton--Interaction}
\label{SubSection:Coupling-Huu}

\begin{center} 
\begin{fmffile}{Figures/Diagrams/FeynDia103} 
\fmfframe(20,20)(20,20){ 
\begin{fmfgraph*}(75,75) 
\fmfleft{l1}
\fmfright{r1,r2}
\fmf{fermion}{v1,l1}
\fmf{fermion}{r1,v1}
\fmf{scalar}{r2,v1}
\fmflabel{$\bar{d}_{{i \alpha}}$}{l1}
\fmflabel{$e_{{j}}$}{r1}
\fmflabel{$\tilde{u}_{{k \gamma}}$}{r2}
\end{fmfgraph*}} 
\end{fmffile} 
\end{center}  
\begin{eqnarray}
 &&i \delta_{\alpha \gamma} U^{e,*}_{R,{j 5}}  \sum_{b=1}^{3}Z^{U,*}_{k b} \sum_{a=1}^{3}U^{d,*}_{R,{i a}} Y_{d,{a b}}\
 P_L
 \nonumber\\ 
  && \,-i \delta_{\alpha \gamma} \Big(g \sum_{a=1}^{3}Z^{U,*}_{k a} U_{L,{i a}}^{d}  U_{L,{j 4}}^{e}  - \sum_{b=1}^{3}\sum_{a=1}^{3}Y^*_{u,{a b}} Z^{U,*}_{k 3 + a}  U_{L,{i b}}^{d}  U_{L,{j 5}}^{e} \Big)
  P_R.
  \label{leptons}
\end{eqnarray}


\subsection
{\bf Up squark-up quark-neutrino--Interaction}
\label{SubSection:Coupling-Huu}

  \begin{center} 
\begin{fmffile}{Figures/Diagrams/FeynDia109} 
\fmfframe(20,20)(20,20){ 
\begin{fmfgraph*}(75,75) 
\fmfleft{l1}
\fmfright{r1,r2}
\fmf{fermion}{v1,l1}
\fmf{plain}{r1,v1}
\fmf{scalar}{r2,v1}
\fmflabel{$\bar{u}_{{i \alpha}}$}{l1}
\fmflabel{$\nu_{{j}}$}{r1}
\fmflabel{$\tilde{u}_{{k \gamma}}$}{r2}
\end{fmfgraph*}} 
\end{fmffile} 
\end{center}  
\begin{eqnarray}
  &&\frac{i}{3} \delta_{\alpha \gamma} \Big(2 \sqrt{2} g' U^{V,*}_{j 4} \sum_{a=1}^{3}Z^{U,*}_{k 3 + a} U^{u,*}_{R,{i a}}   -3 U^{V,*}_{j 7} \sum_{b=1}^{3}Z^{U,*}_{k b} \sum_{a=1}^{3}U^{u,*}_{R,{i a}} Y_{u,{a b}}   \Big)P_L \nonumber \\  
   &&-\frac{i}{6} \delta_{\alpha \gamma} \Big[6 \sum_{b=1}^{3}\sum_{a=1}^{3}Y^*_{u,{a b}} Z^{U,*}_{k 3 + a}  U_{L,{i b}}^{u}  U_{{j 7}}^{V}  + \sqrt{2} \sum_{a=1}^{3}Z^{U,*}_{k a} U_{L,{i a}}^{u}  \Big(3 g U_{{j 5}}^{V}  + g' U_{{j 4}}^{V} \Big)\Big]P_R.\nonumber\\
  \label{neutrinos}
  \end{eqnarray}

\bibliographystyle{utphys}
\bibliography{stop-lsp_munuSSM}

\end{document}